\documentclass[pre,floatfix,twocolumn, 10pt]{revtex4} 
\usepackage{graphicx}
\usepackage{amssymb}
\usepackage{natbib}
\usepackage{dcolumn}
\usepackage{bm}
\usepackage{color}
\usepackage[colorlinks=true, linkcolor=blue, citecolor=blue]{hyperref}
\usepackage{subfigure}
\usepackage{graphicx,psfrag,xspace}
\usepackage{color}
\usepackage{amssymb,amsfonts,amsmath}
\usepackage{setspace}

\begin{document}

\author{E. A. Jagla} 
\affiliation{Centro At\'omico Bariloche, Instituto Balseiro, 
Comisi\'on Nacional de Energ\'ia At\'omica, CNEA, CONICET, UNCUYO,\\
Av.~E.~Bustillo 9500 (R8402AGP) San Carlos de Bariloche, R\'io Negro, Argentina}

\title{Discontinuous depinning/yielding transition of elastic manifolds with tailored internal elasticity}

\begin{abstract} 

We consider elastic manifolds evolving on disordered energy potentials under the action of an external uniform driving. This scenario includes the cases of {\em depinning} and {\em yielding}, 
which provide paradigmatic examples of out of equilibrium phase transitions. In 
both cases, velocity of the manifold is zero at low driving force, and increases smoothly when a critical driving is exceeded,
defining a continuous flow-curve for these systems.
We show that when more general forms of the manifold elasticity are considered, the 
flow curve may become reentrant, and the transition hysteretic, or discontinuous. This constitutes a novel scenario for a discontinuous transition out of equilibrium.
 
\end{abstract}

\maketitle

One paradigmatic example of out-of-equilibrium phase transitions is provided by the phenomenon of depinning of elastic manifolds \cite{fisher,kardar,brazovskii}.  
This can be briefly described in the following terms:
An elastic $d$ dimensional manifold defined by its coordinates $u_i$ ($i\equiv [x_1,...,x_d]$) is considered.
Elastic interactions between different $i$, $j$ are characterized by an interaction kernel $G_{ij}$, in such a way that an increase $du_i$ of $u_i$, generates on site $j$ an increase of force given by $df_j=\sum_j G_{ij}du_i$ (in the simplest case $G_{ij}$ depends only on the distance between sites $i$ and $j$).
If we imagine the elastic interactions being generated by springs of stiffness $k_{ij}$ joining sites $i$ and $j$ ($i\ne j$), then $G_{ij}=k_{ij}$ if $i\ne j$ ($G_{ij}=G_{ji}$), and $G_{ii}=-\sum_j k_{ij}$.
The manifold defined by the coordinates
$u_i$ moves over a set of quenched disordered potentials $V_i(u_i)$ (that we assume to be uncorrelated between different sites) under the action of an external uniform force $F$ applied to all sites. 
In depinning $G_{ij}\geq 0$ ($i\ne j$) (i.e., $k_{ij}\geq 0$). In this case, if the full manifold is connected (i.e., there are no ``islands" of sites with zero interactions to the rest), there is a well defined critical force $F_c$, and the average system velocity $v$ is zero for $F<F_c$, and $v\sim (F-F_c)^\beta$ for $F>F_c$. Typically $\beta\leq 1$, with the limiting case  $\beta=1$ occurring for ``infinite range" interactions. 

A second case of an out of equilibrium transition is that referred to as {\em yielding}, which is typically used to describe the deformation of amorphous materials possessing a finite 
{\em yield stress}\cite{coussot,berthier,nicolas}.
Although in the beginning yielding was studied as a transition rather unrelated
to depinning, it has become clear that, at least in some implementations\cite{jagla1,ferrero2,ferrero3}, the difference between the two cases is actually minor. In fact,
yielding fits in the previous scheme for depinning by simply choosing the kernel $G_{ij}$ appropriately. 
The standard form of $G_{ij}$ for yielding is known as the the Eshelby interaction\cite{eshelby}, which appears when considering the effect of a local volume conserving deformation within an infinite elastic matrix.
Its form is better presented in Fourier space (here for the $d=2$ case, and for one particular symmetry of the deformation) 
\begin{equation}
G_{\bf q}\sim \frac{(q_x^2-q_y^2)^2}{(q_x^2+q_y^2)^2}
\label{eshelby}
\end{equation}
The form of $G_{ij}$ in real space is obtained by Fourier inverting this expression, and it  looks like
\begin{equation}
G_{ij}\sim\frac{\cos (4\theta)}{|{\bf r}_i-{\bf r}_j|^2}
\end{equation}
where $\theta$ is the polar angle ($\tan \theta =y/x$). 
The key property of this interaction (and its main qualitative difference with the depinning case) is that it takes negative values for some  $ij$'s.

The existence of negative values of $G$ poses a non-trivial question on the stability of the system. In fact, in order to be stable, the elastic energy of any given configuration, which is 
$E\equiv \sum_{i,j}G_{ij}u_i u_j/2$, must be definite non-negative. In the present case this is fulfilled due to its original definition in Fourier space, since we can write $E=\sum G_{\bf q}|u_{\bf q}|^2$ which in fact is non negative when $G_{\bf q}$ is given by Eq. (\ref{eshelby}).
The flow curve $v$ vs. $F$ of the model with the Eshelby interaction displays (as for depinning) a critical value $F_c$, such that $v=0$ for $F<F_c$.
In all cases studied in the literature of which we are aware, $v$ increases continuously when $F$ becomes larger that $F_c$,  actually with a power law increase with an exponent $\beta$, which is now larger that one (typically between 1.5 and 2 \cite{ferrero2}).



Beyond the paradigmatic depinning and yielding cases just mentioned, 
there has been particular interest in models that display a {\em discontinuous} transition (i.e., an abrupt jump of $v$ from 0 to a finite value at some critical $F$), as they can justify for instance the appearance of {\em shear bands} in the stationary deformation of glassy materials\cite{sb1,sb2,sb3,sb4,sb5}, an issue of much interest also from the applied point of view (for instance in metallic glasses\cite{mg1,mg2,mg3,mg4}).
To obtain a discontinuous (also referred to as ``first order" or ``reentrant") transition, additional ingredients have been typically added to the basic depinning or yielding model as described above. They include the
aging stabilization/strain rejuvenation scenario
\cite{picard,olmsted,divoux,jagla_2007,
coussot_2002a,mujumdar,martens_2012}, the 
flow-concentration coupling mechanism  
\cite{fall_2009,ovarlaz_2006,besseling}
and the inclusion of inertial effects
\cite{inertia1,inertia2,inertia3}, among others.


We show here that these additional ingredients are in principle not necessary 
to obtain a discontinuous transition as it suffices to choose the
$G_{ij}$ kernel appropriately, without additional ingredients. We have found  different  forms of $G_{ij}$ that produce a discontinuous transition, and we will consider in detail two of them. 
First, we consider the case in which $G_{ij}$ is randomly  chosen (with variable sign). In view of previous comments we see that this choice requires additional considerations about the stability of the model, which we address below. This case can be considered to be 
a variation of the Hèbraud-Lequeux (HL) model\cite{hl0}, in which the elastic interaction between any two sites is taken as a random variable of the coordinates, {\em and time}. Our case can be referred to as a {\em quenched}-HL (qHL) model  in which the interactions are spatially random but chosen once and forever. 
As a second case we consider a modified Eshelby interaction that also displays a discontinuous transition. The remarkable characteristic of this case is that when driven at constant velocity in the coexistence region, the system separates into a part that remains stuck, and a part that flows, in the form of a shear band.

\section{Details of the models}

The model consists of a set of variable $u_i$ defined on the sites $i$ of a $d$-dimensional lattice, interacting through a given elastic kernel $G_{ij}$ as previously indicated. 
The variables $u_i$ evolve onto disordered external potentials $V_i(u_i)$ (which are uncorrelated among different sites).
We consider the potentials $V_i(u_i)$ to be of the {\em narrow well} form. This consists of infinitely narrow wells randomly distributed on the $u$ axis, characterized by the force $f_0$ that is necessary to be applied for a particle in the well to escape from it. For simplicity (and since it does not produce any qualitatively new results) the value of $f_0$ is taken uniform across the whole system. 
The force $f_i$ on site $i$ is the sum of the uniform external force $F$ and the elastic force generated by all other particles: $f_i=F+\sum_j G_{ij}u_j$. 
\begin{equation}
f_i=\sum_j G_{ij}u_j+F
\end{equation}
In the narrow well approximation the dynamic equations are replaced by discrete update rules, as follow.
If at some moment a value $f_i>f_0$ is
detected, the corresponding particle jumps to the next well.
The random positions of the well
are determined on the fly: from position $u_i$ particle
jumps out to a new well at position $\widetilde {u_i}=u_i-\log [\mbox {RND}]$, where $[\mbox {RND}]$ is a random variable with a flat distribution between  0 and 1 (we take the mean distance between wells to be 1).

Two different driving protocols will be used: force driven and velocity driven.
The force driven protocol consists in keeping the value of the external force $F$ as constant, and calculating the average velocity. The instantaneous velocity of the system is determined as the average number  of unstable sites $n$ (i.e., 
those with $f_i>f_0$) divided by the total number of sites in the system $N$. The average velocity is the result of averaging the instantaneous velocity on time.
In the velocity driven protocol we set a  required velocity value $v$. This  fixes the number $n$ of unstable sites at each time step as $n=Nv$. Then the value of the external force $F(t)$ is defined as that producing exactly $n$ unstable sites in the current time step. In this protocol the reported value of $F$ is the average of its temporally fluctuating value.

We first consider the case in which 
 $G_{ij}$ are  {\em random interactions} of variable sign. 
Let us be more precise in this respect.
There is already a well known, mean field model of the yielding transition that considers a random elastic interaction, namely the Hèbraud-Lequeux (HL) model\cite{hl0}. 
In the HL model, each time the variable $u_i$ at site $i$ experiences an increase  $du_i$, the forces $df_j$ in all sites $j$ get a contribution that consists of two pieces. 
There is a deterministic contribution: $df_j=a_0$ if $i\ne j$, and $df_i=-(N-1)a_0$. This can be understood as the effect of elastic springs with stiffness $a_0$ joining any pair of sites.
In addition, there is a random contribution: $df_j= \eta_j$. The values of $\eta_j$ are taken from a zero mean Gaussian distribution\cite{eta_j_medio}. The values of $\eta$ are renewed when another (or the same) variable $u_{i'}$ gets a position increase 
$du_{i'}$. 
These simple rules make the model analytically solvable. The outcome is that there is a continuous yielding transition with a $\beta=2$ exponent.

Our proposal is to consider a case in which the elastic interactions are random in space, but {\em quenched}, namely chosen at the beginning and valid for the full simulation. This choice requires to face a non-trivial issue. The values of $G_{ij}$ have to produce a  definite non-negative energy in order for the model to be stable \cite{hl}. This means that all eigenvalues of the real symmetric $G_{ij}$ matrix must be non-negative.
We enforce this by defining random positive eigenvalues of $G_{ij}$ in some particular base (we use a one-dimensional Fourier base) and transforming back to real space.

With this prescription we obtain values of $G_{ij}$ with the following characteristics.
Each $G_{ij}$ ($i\ne j$) is a random variable of width $\sigma$ with almost zero mean (see below).
Beyond this description of the statistics of individual values of $G_{ij}$, the precise values have subtle correlations that reflect the fact that $G_{\bf q}\geq 0$.
In particular $\sum_j G_{ij}=0$.

\begin{figure}
\includegraphics[width=8cm,clip=true]{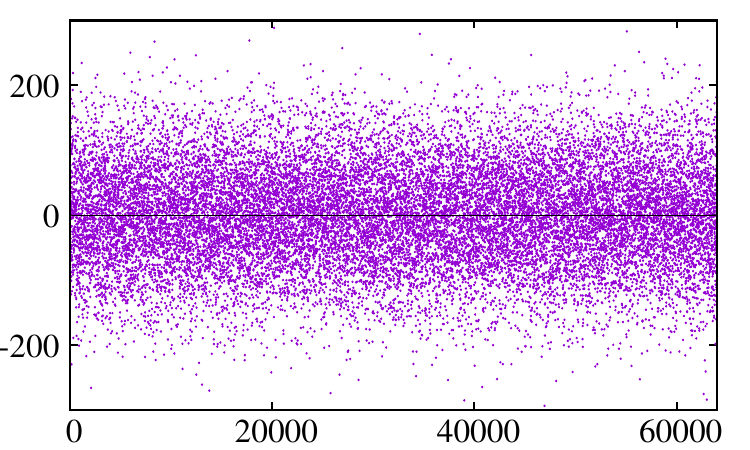}
\caption{The values of $G_{ij}$ as a function of 
$|i-j|$, for a system of $N=2^{16}$, and $p=0.1$. The value of $G_{ij}$ for $i=j$ is 
$\sim -6500$, well out of the scale of the figure. 
The eye does not find any correlation among the data, although 
$G_{q}$ is either 0 or 1.}
\label{gij}
\end{figure}

We consider formally a one-dimensional system of $N$ sites, and define the values of $G_q$ as follows\cite{cos}. If $q\ne 0$ we take
\begin{eqnarray}
G_q&=&1~~~~\mbox {with probability}~p\\
G_q&=&0~~~~\mbox {with probability}~1-p
\end{eqnarray}
and $G_{q=0}=0$. When transforming back to the real space,
we obtain $G_{ii}=-pN$, and $G_{ij}$ (for $i\ne j$) as the sum of $pN$ uncorrelated 
variables of the kind $\cos \eta$, with $\eta$ a random variable. Therefore,  
if $pN \gg 1$,  $G_{ij}$ becomes a Gaussian variable of zero mean, and width $\sim \sqrt{pN}$.
There will be of course correlations between the values of $G_{ij}$ for different $|i-j|$, but
a first glance examination of the values of $G_{ij}$ does not reveal any obvious pattern 
(see Fig. \ref{gij}).

\begin{figure}
\includegraphics[width=8cm,clip=true]{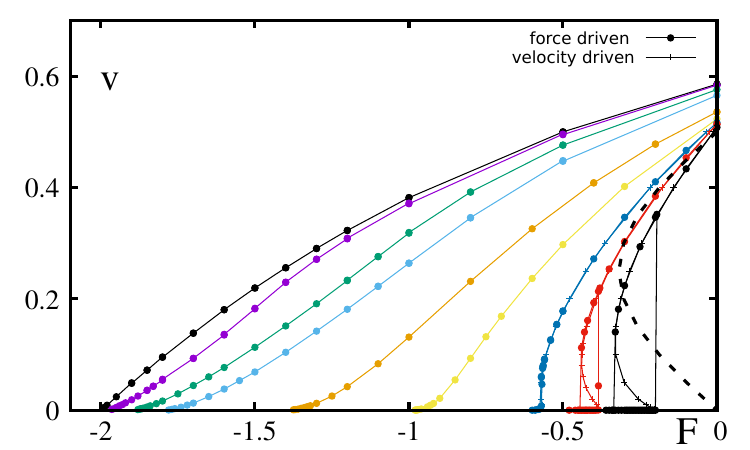}
\caption{Flow curves of the system for different values of $p$ ($p=1, .99, .95, .9 ,.7, .5, .3, .2, .1$, from left to right). Full symbols were obtained with the force driven protocol. The curves display hysteresis for the lowest values of $p$. The crosses show results obtained 
using the velocity driven protocol, and the reentrance is clearly seen. The dashed line is the analytical result for the single-mode model of Section \ref{minimal}.}
\label{flow_curves}
\end{figure}

\section{Results for the quenched Hèbraud-Lequeux model}

In the limit $p= 1$ the form of the $G$ function reduces to $G_{ii}=N$, $G_{ij}=-1/(N-1)$ ($i\ne j$). This corresponds to an infinite range interaction.
As $p$ is reduced, the stochastic part of the interaction increases relatively to the deterministic part. When $p\to 0$ (but always requiring $pN\gg 1$) the interaction is almost purely random, and the local value $G_{ii}$ is not much larger than the typical $G_{ij}$.
The flow curves obtained for a system with $N=2^{18}$ at different values of $p$ are shown in Fig. \ref{flow_curves}.
For $p=1$ we observe the typical mean field behavior with $\beta=1$. As $p$ is reduced, the flow curves get displaced to the right, and there is an apparent increase of the $\beta$ value.
More importantly, there is a remarkable feature of the $v$ vs $F$ flow curves which actually motivated our interest in this model: flow curves become re-entrant when $p$ is sufficiently small, as seen in Fig. \ref{flow_curves} for the lowest values of $p$. This means that the value of $F$ {\em decreases} from $F_c$ when $v$ is increased from zero. This behavior is observed in the case of driving the system by fixing $v$. If the constant force protocol is used instead, we obtain a hysteretic behavior.

We emphasize that the discontinuous transition is obtained in the absence of any other internal mechanism in the model, i.e., no other internal time scales in the system exist.
We remark also that in the present case the reentrant behavior does not imply a spatial coexistence of sites remaining stuck (i.e., with $v=0$) and others flowing with finite velocity. This separation does not occur because we are dealing with a model with {\em long range} interactions in which any interface separating a flowing and a non-flowing region has an energy that increases in time, therefore precluding this separation to occur in the long run. This will be different in the model presented in  section III. In order to gain some insight into the origin of the reentrant transition, we present now a limiting analytical case that displays this behavior.

\subsection{Minimal model with a reentrant transition}
\label{minimal}

The finding of a reentrant (if velocity driven) or discontinuous (if force driven)
flow curve by choosing appropriately the elastic kernel is a novel and surprising result not previously reported. In this respect it is useful to have the simplest possible realization of this scenario, to help in rationalizing the qualitative origin of the phenomenon.

The simplest case for which we have detected a reentrant transition occurs when the elastic kernel is different from zero only at a single value of the wave vector, say $q_0$,
namely $G_q=\delta_{q,q_0}$.
Assuming a one-dimensional geometry, the kernel in real space can thus be written as

\begin{equation}
G_{ij}=-\cos(q_0(i-j))
\label{1modo}
\end{equation}
(in a finite system of $N$ sites with periodic boundary conditions the value $q_0$ must be chosen commensurate, i.e., $q_0 N=2\pi m$ for some integer $m$).
This interaction is clearly non-positive definite. 
With this interaction, the elastic force on site $i$ is given simply by
\begin{equation}
f_i=-\sum_j\cos(q_0(i-j))u_j=-C \cos (q_0i+\phi)
\end{equation}
where $C>0$ and $\phi$ are the amplitude and phase of the $q_0$-Fourier coefficient of the interface profile.
The simple harmonic form of $f_i$ allows to solve analytically for the form of the flow curve.
Consider a velocity driven situation in which we want to drive the interface at some velocity $v$.
This means that at every time step, $Nv$ sites must receive a force larger than the threshold force $f_0$, namely, the condition
\begin{equation}
-C \cos (q_0i+\phi)+F>f_0
\end{equation}
Must be satisfied for $Nv$ out of the total $N$ sites in the system.
 This requires
\begin{equation}
F=f_0+C\cos(\pi v)
\label{sig}
\end{equation}
(note that this value turns out to be independent of the value of $q_0$).
To find the flow curve $v$ vs. $\sigma$ we need to calculate the value of $C$. 
This quantity (as well as $F$) may change along the iterations.
Given a value of $C$ (and, according to (\ref{sig}), a value of $F$) at a given iteration, the value in the next step can be obtained as follows. At velocity $v$, a number $vN$ of the sites in the system will be destabilized at every time step. Due to the simple harmonic form of the force $f_i$ across the system all unstable sites group together near the region where $f_i$ is peaked. These sites change its coordinates to $u_i\to u_i +\delta u_i$, and this produces a change of the Fourier coefficient $C$, that is modified as \cite{foot1}
\begin{equation}
C\to \left |C-\frac 2\pi\sin(\pi v)\right |
\label{c}
\end{equation}
(the absolute value is necessary as the argument is typically alternating in sign).
Note that this is a closed iterative expression to obtain the value of $C$, given a starting value $C_0$.
The iteration produces an alternating sequence of the form $C_0$, $(2/\pi)\sin(\pi v)-C_0$, $C_0$, $(2/\pi)\sin(\pi v)-C_0$, ..., so the average value of $C$ is $\pi^{-1}\sin(\pi v)$, which inserted into \ref{sig} provides the average value of $F$ as
\begin{equation}
F=f_0+\frac 1 \pi\cos(\pi v)\sin(\pi v)
\end{equation}
This is the analytical form of the flow curve we were looking for. It is added to the plot of the qHL model (Fig. \ref{flow_curves}).
It displays clearly a reentrant behavior when $v_0<0.25$ which looks qualitatively similar to the behavior of the full qHL model as $p$ goes to zero.

\begin{figure}[h]
\includegraphics[width=8cm,clip=true]{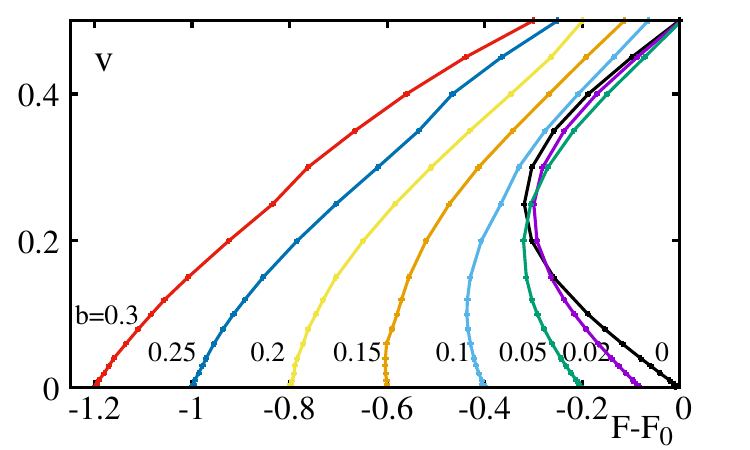}
\caption{Results for a kernel combining a full range mean field interaction of intensity $b$, as indicated, and a single mode interaction of intensity $a=1$.}
\label{1mode+mf}
\end{figure}

We can complement this result by adding a full range (mean field) interaction and considering the  kernel
\begin{equation}
G_{ij}=-a\cos(q_0(i-j))+b
\end{equation}
for $i\ne j$, and 
\begin{equation}
G_{ii}=-a+b(N-1)
\end{equation}
When $a=0$ and $b\ne 0$ we have a pure mean field interaction that is very well understood in its behavior. The flow curve in this case is monotonous, with a $\beta=1$ value. For both $a$, $b \ne 0$ we present here results of numerical simulations that show neatly a transition between a continuous depinning case, when $b$ dominates, to the reentrant case when $a$ dominates. The results are shown in Fig. \ref{1mode+mf}.

\section{Other kernels displaying a discontinuous transition}

\begin{figure}
\includegraphics[width=8cm,clip=true]{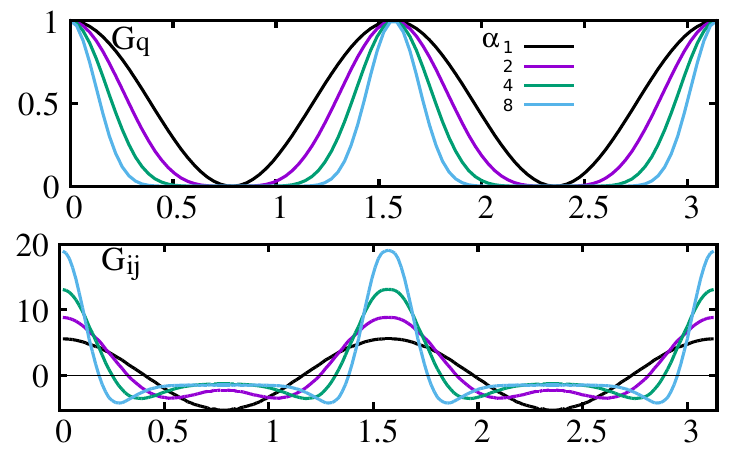}
\caption{Angular dependence of the $\alpha$-Eshelby kernels, in Fourier space (a) and real space (b) for a few values of $\alpha$. } 
\label{kernel_alfa}
\end{figure}

\begin{figure}
\includegraphics[width=8cm,clip=true]{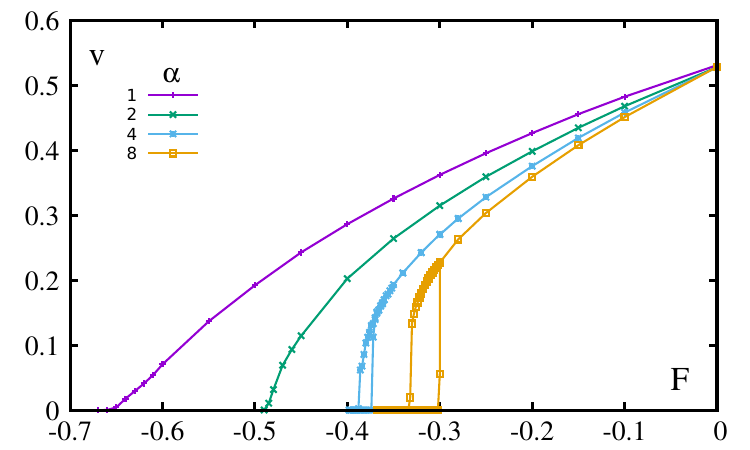}
\caption{Flow curves for kernels $G_\alpha$, as defined in the text. For $\alpha\gtrsim 4$ the flow curves are re-entrant.} 
\label{flow_eshelby}
\end{figure}

\begin{figure}
\includegraphics[width=9cm,clip=true]{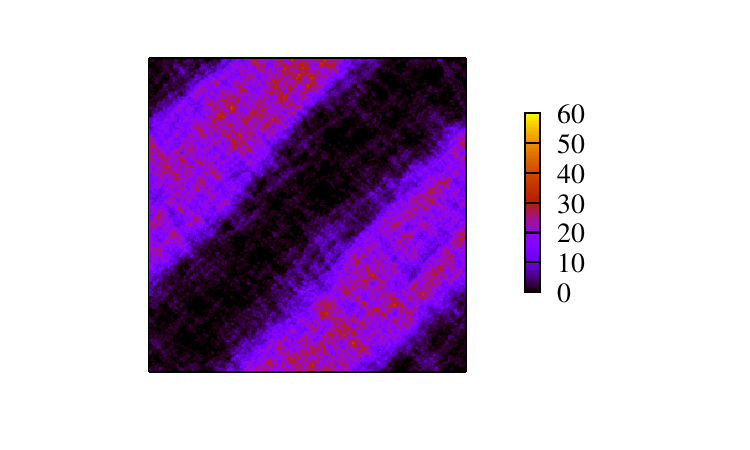}
\caption{Accumulated local deformation in a simulation with $\alpha=8$, at $v=0.05$, in a time interval corresponding to an average deformation of 10. System size is $1024\times 1024$ . A shear band where the system is deforming is clearly visible. } 
\label{shear_band}
\end{figure}

In addition to the quenched random, and single mode kernels of the previous section, we have identified other cases that produce a discontinuous yielding transition in the system. Of particular interest is the case of a kernel obtained by smoothly transforming the Eshelby kernel, in the way that is described now. 
The starting point is the Eshelby kernel in ${\bf q}$-space, as defined in Eq. (\ref{eshelby}).
From this, we obtain a modified kernel dependent of a parameter $\alpha$, through
\begin{equation}
G_{{\bf q},\alpha}\equiv (G_{\bf q})^{\alpha}
\end{equation}
($\alpha>1$). 
The effect of $\alpha$ is to produce narrower lobes in which $G_{{\bf q},\alpha}\sim 1$, and wider regions 
in which $G_{{\bf q},\alpha}\sim 0$. 
The kernel in real space is obtained by back-transforming Fourier $G_{{\bf q},\alpha}$. 
Since $G_{{\bf q},\alpha}$ continues to be independent of the
 modulus of ${\bf q}$, $G_{ij,\alpha}$ still has a decay $\sim 1/|r_i-r_j|^2$.  
 The modification is in 
 the angular dependence of the kernel. In Fig. \ref{kernel_alfa} we see the effect of $\alpha$ on the 
 angular dependence of $G_{\bf q,\alpha}$, and $G_{ij,\alpha }$. Simulations using kernels with 
 different values of $\alpha$ provide the flow curves shown in Fig. \ref{flow_eshelby}. A discontinuous transition is obtained for $\alpha \gtrsim 4$ \cite{orientacion}. At variance with the results for the random kernel in the previous section, now the reentrant transition produces (in the velocity driven protocol) a true spatial separation of the system in flowing and a non-flowing parts. This can be seen in Fig. \ref{shear_band}, where we plotted the spatial distribution of the deformation of the system, in a velocity driven simulation at $v=0.05$ and $\alpha=8$. A  shear band oriented  along one diagonal is clearly visible. This band can appear since  the directions $q_x=\pm q_y$ correspond  to zero-energy modes, as in the original Eshelby kernel. Therefore the existence of a shear band like that seen in Fig. \ref{shear_band} does not create any energy excess.

\section{Summary and conclusions}

We have shown that standard models for depinning or yielding can be tailored to produce discontinuous transitions by appropriately choosing the form of the elastic energy of the system. Three cases have been presented: a {\em random} elastic interaction of alternating sign, a single mode interaction ($\sim \cos(q_0|i-j|)$) in a one dimensional case, and a modified Eshelby interaction in a two-dimensional system. No additional ingredients like aging mechanisms or inertial effects are neccessary to obtain this behavior. This is not a claim that reentrant transitions found in particular experimental situations are due to the existence of exotic elatic interactions. Our claims are a proof of concept that this could eventually happen in systems with some exotic elastic interactions. 
For instance, in Ref. (\cite{rosso}) the effect of difference in shear modulus of a material in two different directions rotated $\pi/4$ from each other was analysed. The structure of the Eshelby interaction was seen to be modified in a way that resembles our $\alpha$-kernels. Yet, in the case of Ref. (\cite{rosso}) the symmetry between possitive and negative values of $G_{ij}$ was maintained, and no sign of a reentrant behavior was observed. Our results are a motivation to search for other kind of non-symetric interactions that may give rise to a discontinuous 
yielding transition and a robust shear banding phenomenon.




\begin{thebibliography}{1}
\bibitem{eta_j_medio} Typically, an additional small constant term is added to guarantee that $\sum_j df_j=0$.
\bibitem{hl}In the HL this is not a requirement. The permanent renewal of the values of $G$ stabilizes the system, and renders this prescription unnecessary.
\bibitem{lineal} The value of $F_c$ for each value of $p$ is determined so as to obtain the best power law fitting of the flow curve close to $F_c$. Incidentally, we notice that the values of $F_c$ obtained follow very accurately a linear trend with $p$, of the form $F_c=F_0-2p$. Although we have not been able to prove it analytically, this simple analytical dependence may call for a more detailed analytical investigation.
\bibitem{foot1} The values of $\delta u_i$ are exponentially distributed variables. For the calculation of the new value of $C$ we take account that many neighbor sites are destabilized, and therefore the 
modification if the value of $C$ can be calculated considering only the average value of $\delta u_i$, which is 1.
\bibitem{orientacion} the minimum value of $\alpha$ to obtain a discontinuous transition is dependent on the spatial symmetry of the Eshelby kernel used. In case we take the starting kernel as $G_{\bf q}\sim 4q_x^2q_y^2/(q_x^2+q_y^2)^2$ (which is obtained a s a $\pi/4$ rotation from the one used in the text) the minimum $\alpha$ is larger, near 10. This dependence should not be surprising as the long range nature of the kernel interacts with the symmetry of the simulation box, which is square.
\end{thebibliography}
\end{document}